\documentclass[12pt]{article}
\usepackage{epsfig,epsf,amsmath}

\newlength{\dinwidth}
\newlength{\dinmargin}
\def\lapproxeq{\lower .7ex\hbox{$\;\stackrel{\textstyle <}{\sim}\;$}}
\def\gapproxeq{\lower .7ex\hbox{$\;\stackrel{\textstyle >}{\sim}\;$}}

\def\be{\begin{equation}}
\def\ee{\end{equation}}
\def\bea{\begin{eqnarray}}
\def\eea{\end{eqnarray}}

 \def\hpz{\hphantom{0}} \def\hpzz{\hphantom{00}}

\begin{document}
\titlepage

\begin{flushright}
LTH 954\\
TU--916\\
21 August 2012\\
\end{flushright}

\vspace*{4cm}

\begin{center}
{\Large \bf Hadronic contributions to the anomalous magnetic moment of
the electron and the hyperfine splitting of muonium}

\vspace*{1cm} {\sc Daisuke Nomura}$^a$ and {\sc Thomas Teubner}$^{b}$

\vspace*{0.5cm}
$^a$ {\em Department of Physics, Tohoku University, Sendai 980-8578, Japan}\\
$^b$ {\em Department of Mathematical Sciences, University of Liverpool, Liverpool L69 3BX, U.K.}\\
\end{center}

\vspace*{1cm}

\begin{abstract}
\noindent
Motivated by recent progress of theory and experiment
on the anomalous magnetic moment of the electron, $a_e$, we
update the hadronic contributions to $a_e$.
Using our up-to-date compilation of $e^+ e^-\to {\rm hadrons}$
data, we find the leading order hadronic
contribution $a_e^{\rm had,\,LO,\,VP} 
= (1.866 \pm 0.010_{\rm exp} \pm 0.005_{\rm rad} )\cdot 10^{-12}$ and 
the next-to-leading order hadronic contribution 
$a_e^{\rm had,\,NLO,\,VP} = (- 0.2234 \pm 0.0012_{\rm exp} \pm
0.0007_{\rm rad} ) \cdot 10^{-12} \,$,
where the first and second errors are from the error of 
the experimental data and the uncertainty in the treatment of
radiative corrections, respectively.  These values are compatible with
earlier evaluations by other groups, but have significantly
improved uncertainties due to the more precise input data used.
We also update the leading order hadronic contribution
to the ground state hyperfine splitting of muonium, obtaining
$\Delta \nu_{\rm Mu}^{\rm had,\,VP} = 
(232.68 \pm 1.25_{\rm exp} \pm 0.72_{\rm rad})$Hz.
This value is consistent with the most precise evaluation 
in the literature and reduces its error by a factor of two.
\end{abstract}

\section{Introduction}
With the start of the LHC, the hunt for physics beyond the Standard
Model (SM) at the energy frontier has entered a new era, though so far
no direct signal for `new physics' has been observed. At the same
time, experiments at lower energies are measuring SM parameters with
unprecedented accuracy and are becoming ever more sensitive to
quantum effects, possibly from physics beyond the SM. A prime example
for this is the anomalous magnetic moment of leptons
\cite{Roberts:2010zz}, caused by loop effects mainly from
Quantum Electro-Dynamics (QED) but also influenced by the strong and
weak force and possibly by effects from new physics. For the muon, the
anomaly $a_\mu$ is sensitive to all sectors of the SM and has been
measured with 0.5 ppm accuracy, using a muon storage ring at
Brookhaven \cite{BNL}. This measurement, when compared to the SM
prediction of $a_\mu$, shows a discrepancy of $3.3\,\sigma$
\cite{Gm2Reviews,HLMNT,Davier:2010nc} which may be a sign for new
physics.\footnote{The tau lepton's anomaly, $a_\tau$, is even more
  sensitive to physics at higher energy scales than $a_\mu$, but is
  very difficult to measure with high precision due to the short
  lifetime of the tau. See \cite{Eidelman:2007sb} for a recent review
  and SM prediction of $a_\tau$.} 

The electron's anomalous magnetic moment, $a_e$, has been measured at
Harvard via atomic spectroscopy \cite{Harvard}, 
\begin{equation}
a_e = 1159~652~180.73~(28) \cdot 10^{-12}\,.
\label{eq:Harvard}
\end{equation}
This corresponds to a $0.24$ ppb accuracy, hence $a_e$ is known even
much more precisely than $a_\mu$. However, due to the small mass of
the electron, $a_e$ is less sensitive to quantum effects from higher
mass scales and is mainly generated by QED effects, though it also
receives small contributions from the strong and weak interaction. It
is therefore an ideal place to determine the fine structure constant
$\alpha$. 

Another precision observable is the ground state hyperfine
splitting (HFS) of muonium. Similar to $a_e$, it is sensitive
to quantum effects at low
mass scales and mainly QED dominated. It is very useful to determine
the electron-to-muon mass ratio and hence the muon mass. 

To determine SM parameters or to become sensitive to new physics, it
is crucial that the theoretical predictions of such precision
observables have accuracies similar to or smaller than the
experimental uncertainties. For the lepton anomalies and also the HFS
of muonium this has been achieved by higher-order calculations in all
sectors of the SM (and various extensions of it)
\cite{Gm2Reviews,Jegerlehner:2009ry,Eides:2000xc}. For $a_e$, the
group of Kinoshita and collaborators have, after a many-year effort,
completed calculations including full five-loop effects
\cite{Aoyama:2012wj,Kinoshitaetal-fourthorder,Kinoshita:2005sm,
  Aoyama:2008gy,Aoyama:2008hz,Aoyama:2010yt,Aoyama_recent}. These
results have reduced the error of the 
theoretical prediction of $a_e$ by more than four-fold. It is
therefore timely to provide an up-to-date evaluation of the hadronic
contributions. In this short letter we calculate the hadronic
contributions due to vacuum polarisation effects for the anomalous
magnetic moment of the electron and also for the hyperfine
splitting of the ground state of muonium.

\section{Anomalous magnetic moment of the electron}
At present, the most precise determination of the fine structure
constant is through the measurement of the anomalous magnetic moment
of the electron, $a_e$, using the Harvard measurement \cite{Harvard}
quoted in Eq.~(\ref{eq:Harvard}) above. With their new complete
five-loop and updated and improved four-loop QED contributions, and
using results from \cite{DavierHocker,Krause,aeHadlbyl,aeEW} for the
hadronic and electroweak (EW) contributions, Kinoshita {\em et al.}
arrive at \cite{Aoyama:2012wj}
\begin{equation}
\alpha^{-1}(a_e) = 137.035~999~1736~(68)(46)(26)(331)\,,
\label{eq:alpha-ae}
\end{equation} 
corresponding to a 0.25~ppb accuracy. This value is slightly
different but compatible (at the level of $1.3\,\sigma$) with the
second best determination of $\alpha$ which is using Rubidium atoms
\cite{Rb11,Mohr:2012tt} and reads $\alpha^{-1}({\rm Rb})
= 137.035~999~049~(90)$ [0.66~ppb].\footnote{Note that this value is
  a big improvement from the previous measurement by the same group
  \cite{Rb08}.} 
The four errors quoted in (\ref{eq:alpha-ae}) stem from the four- and
five-loop QED contributions, from the hadronic and electroweak
effects, and from the uncertainty of the experimental value for $a_e$.
The group of Gabrielse is working on further improvements of the
measurement of $a_e$ and also a similar determination of the positron's
anomaly which, assuming CPT invariance, can be combined with the electron's
anomaly similar to the case of the muon. However, currently the
experimental error of $a_e$ is dominating the theoretical
uncertainties, which have been reduced dramatically by the new QED
calculations of Kinoshita and collaborators. Their four- and five-loop
errors, i.e.\ the first and second error in (\ref{eq:alpha-ae}), are
mainly coming from the statistical uncertainties of the numerical
Monte Carlo integrations and can be improved with increased
computational efforts when needed. To understand the composition of
the third error, let us list the hadronic and electroweak
contributions as used in~\cite{Aoyama:2012wj}: 
\begin{eqnarray}
a_e^{\rm had,\,LO,\,VP} &=& 1.875~(18) \cdot 10^{-12}\,,
\label{eq:ae-hadLOVP-prev} \\
a_e^{\rm had,\,NLO,\,VP} &=& -0.225~(5) \cdot 10^{-12}\,,
\label{eq:ae-had-NLO-Krause} \\
a_e^{\rm had,\,l\mbox{-}by\mbox{-}l}&=& 0.035~(10) \cdot 10^{-12}\,,
\label{eq:ae-lbyl} \\
a_e^{\rm EW} &=& 0.0297~(5) \cdot 10^{-12}\,.
\label{eq:ae-had-ew}
\end{eqnarray}
Here, similar to the case of the anomalous magnetic moment of the
muon, the error from the electroweak effects is negligible compared to
the uncertainties from the hadronic contributions. The latter are
dominated by the leading order (LO) and next-to-leading order (NLO)
hadronic vacuum polarisation (VP) corrections.
For $a_e$, the contributions from light-by-light (l-by-l) scattering
diagrams are rather small, but they still add significantly to
the uncertainty of the $a_e^{\rm had}$. A further scrutiny of the
model calculations, and possibly also `first-principle'
determinations based on lattice calculations, are under way for the
anomalous magnetic moment of the muon. This in turn will also allow to
improve the estimates for $a_e^{\rm had,\,l\mbox{-}by\mbox{-}l}$.
In the following,
we will discuss our up-to-date determination of $a_e^{\rm had,\,LO,\,VP}$
and $a_e^{\rm had,\,NLO,\,VP}$. Note that while the hadronic cross
section data used as input in these calculations have improved
significantly in recent years due to both direct scan measurements and
via the method of radiative return (see e.g.\ \cite{MCSIGHAD} for a
recent review of the field), no up-to-date determination of the
VP induced contributions is available so far.

\subsection{Leading order hadronic VP contributions}
The calculation of the leading order hadronic VP contributions to
$a_e$ is very similar to the case of the muon and uses a dispersion
integral over the hadronic cross section times a well-known kernel
function $K_\ell(s)$, see e.g.\ the detailed discussion in \cite{HMNT}:
\begin{align}
  a^{\rm had,\,LO,\,VP}_\ell
= 
  \frac{1}{4\pi^3} \int_{m_{\pi^0}^2}^\infty {\rm d}s ~ K_{\ell}(s)
        \sigma^0_{\rm had}(s)\,,
\end{align}
where $\ell = e, \mu$, and $\sigma^0_{\rm had}(s)$ is the
undressed (i.e.\ excluding VP corrections) total hadronic
cross section.
If we define the function $\hat{K}_\ell(s)$ by
\begin{eqnarray}
 \hat{K}_\ell(s) = \frac{3s}{m_\ell^2} K_\ell(s)\,,
\end{eqnarray}
then $\hat{K}_\ell(s)$ is a monotonically increasing function of
order one with $\hat{K}_\ell(s) \to 1$ for $s \to\infty$. 
In the case of $a_e$, 
the kernel $\hat{K}_e(s)$ is very close to one throughout.  
In fact, $\hat{K}_e(s)$ can be expanded in terms of $m_e^2/s$ as
\begin{align}
 \hat{K}_e(s) = 1 + \left( 3 \ln \frac{m_e^2}{s} + \frac{25}{4} \right)
 \frac{m_e^2}{s} 
+ {\cal O}\left( \frac{m_e^4}{s^2} \ln \frac{m_e^2}{s}\right)\,,
\end{align}
and the deviation of $\hat{K}_e(s)$ from one is almost always
negligible.  (In the case of $a_\mu$, $\hat{K}_\mu(s=m_{\pi^0}^2)=0.40$, 
$\hat{K}_\mu(s=4m_{\pi^\pm}^2)=0.63$, and 
$\hat{K}_\mu(s) \to 1$ for $s \to\infty$.)
Hence the low energy contributions to the
dispersion integral are even more important than in the case of the
muon. Using our up-to-date comprehensive compilation of hadronic cross
section data \cite{HLMNT} we obtain contributions to $a_e$ from
different energy regions as displayed in Tab.~\ref{table-ae-hfs}.
\begin{table}[tb]
\begin{center}
 \begin{tabular}{l|c|c}
 \hline
 \rule{0pt}{2.5ex}
 Energy region, remarks & contribution to $a_e$ 
                  & contribution to $\Delta \nu_{\rm Mu}^{\rm had,\,VP}$
                          \\ \hline \hline
 $2m_{\pi}-0.305\,{\rm GeV}\,({\rm ChPT},\,2\pi)$ &
                  $ 0.0031 \pm 0.0001$
                & $ \hpzz0.25 \pm 0.01_{\phantom{\rm exp}} $ \\ 
 $3m_{\pi}-0.66\,{\rm GeV}\,({\rm ChPT},\,3\pi)$ 
                & $ 0 $ & $ \hpzz0_{\phantom{\rm exp}} $ \\
 $m_{\pi}-0.60\,{\rm GeV}\,({\rm ChPT},\,\pi^0\gamma)$ 
                & $ 0.0004 \pm 0.0000 $
                & $ \hpzz0.04 \pm 0.00_{\phantom{\rm exp}} $ \\
 $m_{\eta}-0.69\,{\rm GeV}\,({\rm ChPT},\,\eta\gamma)$
                & $ 0 $ & $ \hpzz0_{\phantom{\rm exp}} $ \\
 $\phi\rightarrow\,$unaccounted modes 
                & $ 0.0001 \pm 0.0001 $
                & $ \hpzz0.01 \pm 0.01_{\phantom{\rm exp}} $ \\ \hline
 $0.305 - 1.43\,{\rm GeV}$ (exclusive channels)
                & $ 1.6565 \pm 0.0092 $
                & $ 195.15 \pm 1.07_{\phantom{\rm exp}} $ \\
 $1.43 - 2\,{\rm GeV}$ (excl.~data + isospin relations)
                & $ 0.0846 \pm 0.0027 $
                & $ \hpz13.41 \pm 0.43_{\phantom{\rm exp}} $ \\
 $2 - 2.6\,{\rm GeV}$ (inclusive data) 
                & $ 0.0378 \pm 0.0015 $ 
                & $ \hpzz6.54 \pm 0.26_{\phantom{\rm exp}} $ \\
 $2.6 - 3.73\,{\rm GeV}$ (pQCD with BESII error) 
                & $ 0.0257 \pm 0.0009 $
                & $ \hpzz4.84 \pm 0.17_{\phantom{\rm exp}} $ \\
 $3.73 - 11.09\,{\rm GeV}$ (incl.~data)
                & $ 0.0345 \pm 0.0005 $
                & $ \hpzz7.47 \pm 0.10_{\phantom{\rm exp}} $ \\ \hline
 $J/\psi+\psi'$ & $ 0.0185 \pm 0.0004 $
                & $ \hpzz3.53 \pm 0.07_{\phantom{\rm exp}} $ \\
 $\Upsilon(1{\rm S}-6{\rm S})$ 
                & $ 0.0002 \pm 0.0000 $  
                & $ \hpzz0.06 \pm 0.00_{\phantom{\rm exp}} $ \\ \hline
 $11.09 - \infty\,{\rm GeV}$ (pQCD)
                & $ 0.0049 \pm 0.0000 $
                & $ \hpzz1.37  \pm 0.00_{\phantom{\rm exp}} $ \\ \hline
Total & $ 1.866 \pm 0.010_{\rm exp}$ &
        $ 232.68 \pm 1.25_{\rm exp}$ \\ \hline
 \end{tabular}
\end{center}
\vspace{-2mm}
\caption{
Contributions to $a_e^{\rm had,\,LO,\,VP}$ (in units of $10^{-12}$)
and to $\Delta \nu_{\rm Mu}^{\rm had,\,VP}$ (in Hz) from different energy
regions, obtained with the data compilation as used in \cite{HLMNT}.
The first four lines give our predictions of contributions close to
threshold where no data are available and are based on chiral
perturbation theory (ChPT), see \cite{HMNT} for details. For $2.6 <
\sqrt{s} < 3.73$ GeV perturbative QCD (pQCD) with errors comparable to
those of the latest BES data~\cite{Ablikim:2009ad} in this energy
region is used. In the region below $2$ GeV the sum of exclusive
channels, supplemented by isospin relations for channels where no
data are available, is used, see \cite{HLMNT} for details.} 
\label{table-ae-hfs}
\end{table}
The errors given in Tab.~\ref{table-ae-hfs} contain the statistical and
systematic uncertainties from the input data including
correlations over different energies for the various hadronic final
states which are added incoherently. In addition, we have to take into
account the uncertainties in the treatment of radiative corrections from
final state radiation and VP effects, see \cite{HLMNT,HMNT} for a
detailed discussion. These are conservatively estimated to be  
\begin{equation}
\delta a_e^{\rm FSR} = 0.005\,,\qquad \delta a_e^{\rm VP} = 0.002
\label{eq:delaeradcor}
\end{equation}
and have to be added in quadrature to the error of the total $a_e^{\rm had,\,
  LO,\,VP}$ which hence reads
\begin{equation}
a_e^{\rm had,\,LO,\,VP} = (1.866 \pm 0.011) \cdot 10^{-12} \,.
\label{eq:aehadvplototal}
\end{equation}
This number is to be compared with the result from \cite{DavierHocker}
given in Eq.~(\ref{eq:ae-hadLOVP-prev}), which has been used in
\cite{Aoyama:2012wj} and \cite{Mohr:2012tt}. As expected, the mean
value has changed only slightly, but the total error is significantly
improved by about $40\%$ due to the improved hadronic cross section
data. Compared with the more recent evaluation given in
\cite{Jegerlehner:2009ry}, which is $a_e^{\rm had,\,LO,\,VP} =
(1.860 \pm 0.015) \cdot 10^{-12}$, the agreement is very good with a
further but less significant improvement in the error.

\subsection{Next-to-leading order hadronic VP contributions and total
  $a_e^{\rm had}$}
Contrary to the next-to-leading order (NLO) contributions induced by
so-called hadronic light-by-light scattering diagrams, the NLO
corrections including hadronic VP diagrams can be calculated with help
of dispersion integrals, see e.g.~\cite{HLMNT,Krause,HMNT}. Due to the
small mass of the electron, diagrams with two hadronic VP insertions
or one hadronic and one VP insertion from a heavy lepton (muon or tau)
are strongly suppressed, and the diagrams with one hadronic VP
insertion and one additional photon or electron loop are practically
the only relevant contributions. With our latest compilation of hadronic
cross section data we obtain the value
\begin{equation}
a_e^{\rm had,\,NLO,\,VP} = \left(- 0.2234 \pm 0.0012_{\rm exp} \pm
0.0007_{\rm rad} \right)\cdot 10^{-12} \,,
\label{eq:aehadvpnlo}
\end{equation}
where the uncertainty due to the statistical and systematic errors of
the hadronic cross section data used as input and the additional
error due to radiative corrections applied to the data are given separately.
This result is close to the result quoted in Eq.~(\ref{eq:ae-had-NLO-Krause})
\cite{Krause}, but has a much smaller error.
These NLO corrections lead to a reduction of the hadronic LO VP
corrections by about 12\%, so should not be neglected.

Combining the results from
Eqs.~(\ref{eq:aehadvplototal}) and (\ref{eq:aehadvpnlo})\footnote{The LO and
NLO VP contributions are, to a very good approximation, totally
correlated, which is taken into account accordingly in the error
combination.} with $a_e^{\rm had,\,l\mbox{-}by\mbox{-}l}$
from (\ref{eq:ae-lbyl})
\cite{aeHadlbyl}\footnote{This value for the light-by-light
  contributions is compatible with the evaluation by Jegerlehner and
  Nyffeler~\cite{Jegerlehner:2009ry,Nyffeler:2009tw}, who quote
  $a_e^{\rm had,\,l\mbox{-}by\mbox{-}l}= 0.039~(13) \cdot 10^{-12}$.},
our estimate of the hadronic contributions to the electron's anomaly is 
\begin{equation}
a_e^{\rm had} = \left(1.678 \pm 0.014 \right)\cdot 10^{-12} \,.
\label{eq:aehad}
\end{equation}

Equations (\ref{eq:aehadvplototal}) and (\ref{eq:aehadvpnlo}) are the
main result of this letter.
Given the current uncertainties of the QED result for $a_e$ and the
uncertainty of the experimental measurement, the improvements
in the hadronic contributions may not look dramatic and do not affect
the determination of $\alpha$ significantly. However, anticipating
future improvements for $a_e^{\rm QED}$ and $a_e^{\rm exp}$, it is
important that the hadronic contributions are now much better under
control and will not limit the indirect determination of $\alpha$ for
the foreseeable future.

\section{Hyperfine splitting of muonium}
Another high-precision observable is the HFS of the ground state of
muonium. Comparing its measured and predicted values is currently the
best method to determine the electron-to-muon mass ratio and hence the
muon mass~\cite{Mohr:2012tt}. The HFS is mainly QED dominated, but
also receives higher-order contributions due to hadronic and EW
interactions, see
\cite{Eides:2000xc,Mohr:2012tt,EidesGrotchShelyuto-book} for detailed
reviews and further references. 

The experimental value of the muonium HFS
is~\cite{Liu:1999iz,Mariam:1982bq}:
\begin{align}
 \nu_{\rm Mu}({\rm exp}) = 4~463~302~776(51) \text{~Hz}\,,
\label{eq:MuHFS_exp}
\end{align}
while the theoretical value quoted in the 2010 version of
CODATA~\cite{Mohr:2012tt} is
\begin{align}
 \nu_{\rm Mu}(\text{theory}) = 4~463~302~891 (272) \text{~Hz}\,,
\end{align}
where the uncertainty is dominated by that of the mass ratio
$m_e/m_\mu$.  According to Ref.~\cite{Mohr:2012tt},
among the theoretical uncertainty of 272 Hz, the uncertainty
of the QED contributions is 98 Hz and that of the hadronic
contribution 4 Hz.  The hadronic contribution quoted 
in Ref.~\cite{Mohr:2012tt} is
\begin{align}
 \Delta \nu_{\rm Mu}^{\rm had} = 236(4) \text{~Hz}\,,
\label{eq:nuMuhad}
\end{align}
which is the sum of the hadronic VP contribution $\Delta \nu_{\rm
  Mu}^{\rm had,\,VP} = 231.2(2.9)$ Hz~\cite{Eidelman:2002na} and the
hadronic higher-order contribution $\Delta \nu_{\rm Mu}^{\rm
  had,\,HO}= 5(2)$ Hz~\cite{Eidelman:2002na}. Contrary to the case of
the lepton anomalies, for the muonium HFS the hadronic light-by-light
contributions are completely negligible at the current level of
accuracy~\cite{Karshenboim:2008gg}. While the error in the theoretical
prediction of the muonium HFS is dominated by the estimates of unknown
higher-order QED contributions, the hadronic contribution is not
negligible and the error quoted in (\ref{eq:nuMuhad}) is just about
one order of magnitude below the current experimental
error. In fact, there is a planned experiment to measure the muonium
HFS at J-PARC~\cite{J-PARC_MuHFS}, which aims at reducing the
experimental uncertainty by a factor of two or more compared to
Eq.~(\ref{eq:MuHFS_exp}).\footnote{Also note that in their most recent
  work on QED corrections to the HFS of muonium, Eides and Shelyuto
  quote an uncertainty of about 10 Hz as current goal for the
  theoretical uncertainty of the HFS \cite{Eides:2012ea}.}  We
therefore take the opportunity to improve the hadronic contributions.
The hadronic VP contributions to the muonium HFS, $\Delta \nu_{\rm
  Mu}^{\rm had,\,VP}$, have previously been evaluated by a number of
groups~\cite{Eidelman:2002na,Sapirstein:1983xr,Karimkhodzhaev:1991zq,
  Faustov:1998kx,Czarnecki:2001yx,Karshenboim:2001yy,Narison:2001xj}.
They can be written as a dispersion integral~\cite{Sapirstein:1983xr},
\begin{align}
\Delta \nu_{\rm Mu}^{\rm had,\,VP}
 = \frac{1}{2\pi^3} \frac{m_e}{m_\mu} \nu_F
   \int_{m_{\pi^0}^2}^\infty {\rm d}s ~ K_{\rm Mu}(s) \sigma^0_{\rm had}(s),
\end{align}
where $\nu_F$ is the so-called Fermi energy,
\begin{align}
 \nu_F = \frac{16}{3} R_\infty \alpha^2 \frac{m_e}{m_\mu}
         \left[ 1 + \frac{m_e}{m_\mu} \right]^{-3}\,,
\end{align}
where $R_\infty$ is the Rydberg constant,
$R_\infty = 3~289~841~960~364(17)$ kHz~\cite{Mohr:2012tt}.
The explicit form of the kernel function $K_{\rm Mu}(s)$
is given e.g.\ in Ref.~\cite{Karshenboim:2001yy}.  After
correcting typos in Ref.~\cite{Karshenboim:2001yy} it reads
\begin{align}
  K_{\rm Mu}(s) = \begin{cases}
- 2 \left( \frac{s}{4 m_\mu^2} + 2 \right)
 L\left(\sqrt{\frac{s}{4m_\mu^2}}\right)
+ \left( \frac{s}{4 m_\mu^2} + \frac32 \right)
   \ln \frac{s}{m^2_\mu} - \frac12, &
                         (\text{for } s < 4m_\mu^2)\,,\\
 - \left( \frac{s}{4 m_\mu^2} + 2 \right)
     \beta \ln \left( \frac{1 + \beta }{ 1 - \beta } \right)
          + \left( \frac{s}{4 m_\mu^2} + \frac32 \right)
          \ln \frac{s}{m_\mu^2}
          - \frac12, &           
                         (\text{for } s \ge 4m_\mu^2)\,,
\end{cases}
\end{align}
where
\begin{align}
 L(\tau) \equiv - \frac{\sqrt{1-\tau^2}}{\tau}
             \tan^{-1} \frac{\sqrt{1-\tau^2}}{\tau},
~~~~~~~~~  \beta \equiv \sqrt{ 1 - \frac{4 m_\mu^2}{s} } ~ .
\end{align}
The kernel $K_{\rm Mu}(s)$ is a monotonically decreasing function
with $K_{\rm Mu}(s=m_{\pi^0}^2)=5.53$ and 
$K_{\rm Mu}(s=4 m_{\pi^\pm}^2)=1.98$, approaching zero
as $s\to \infty$.
For large $s$, an expansion in terms of $m_\mu^2/s$ is useful:
\begin{align}
K_{\rm Mu}(s) = &
  \left( - \frac92 \ln \frac{m_\mu^2}{s} + \frac{15}4 \right) 
  \frac{m_\mu^2}{s}
+{\cal O}\left( \frac{m_\mu^4}{s^2} \ln \frac{m_\mu^2}{s} \right)~.
\label{eq:KMus-expand}
\end{align}
Basically the dependence of $K_{\rm Mu}(s)$ on $s$ is similar to
that of $K_\mu(s)$.  Since the leading term in the above expansion is
$(m_\mu^2/s)\ln (s/m_\mu^2)$ rather than $(m_\mu^2/s)$,
it puts only slightly more emphasis on higher energies compared to 
the hadronic VP corrections to $a_\mu$.

Using the same input as above, we obtain the hadronic contributions
from different energy regions as listed in Tab.~\ref{table-ae-hfs}.
As in the case of $a_e$ above, the error displayed in the last
line of Tab.~\ref{table-ae-hfs} contains only the uncertainties of the
experimental data. Adding an error from the conservatively estimated
uncertainty due to radiative corrections, $\pm 0.72_{\rm rad}$~Hz, our
final result for the hadronic VP contributions to the muonium HFS
reads
\begin{align}
\Delta \nu_{\rm Mu}^{\rm had,\,VP} = \left(232.68 \pm
  1.44\right)\text{~Hz}\,. 
\label{eq:DeltanuMuhad}
\end{align}
Our result is compatible with and, as expected, considerably more
accurate than the previous result quoted in
Eq.~(\ref{eq:nuMuhad}). Note however, that so far no attempt has been
made at a complete calculation of higher-order hadronic VP corrections
which have been estimated to be of the same order or bigger than the
error of the leading order ones
\cite{Eidelman:2002na,Karshenboim:2001yy}.

\section{Conclusions}
We have used our comprehensive compilation of hadronic cross section
data to determine the hadronic vacuum polarisation contributions to
the anomalous magnetic moment of the electron, $a_e$, to leading and
next-to-leading order. We have also evaluated the hadronic VP contributions
to the hyperfine splitting of the ground state of muonium. Our main results are
given in Eqs.~(\ref{eq:aehadvplototal}), (\ref{eq:aehadvpnlo}) and
(\ref{eq:DeltanuMuhad}). While the changes of the central values
compared to earlier determinations are modest, the corresponding
uncertainties have been significantly improved and will, for the
foreseeable future, not affect precision determinations of physical
constants from these observables.

\vspace{1cm}
\noindent
{\bf\large Acknowledgements}

\vspace{4mm}
\noindent
We thank the Research Centre for Mathematical Modelling in the
Department of Mathematical Sciences for supporting a visit of DN at
Liverpool.  DN thanks M.\ Nio for useful discussions on $a_e$.

\end{document}